# Reversible Tuning of the Heavy Fermion Ground State in CeCoIn$_5$

L. D. Pham[1], Tuson Park[2], S. Maquilon[1], J. D. Thompson[2] and Z. Fisk[3]

[1] *University of California, Davis, CA 95616*

[2] *Los Alamos National Laboratory, Los Alamos, NM 87545*

[3] *University of California, Irvine, CA 92697-4573*

Cadmium-doping the heavy-fermion superconductor CeCoIn$_5$ at the percent level acts as an electronic tuning agent, sensitively shifting the balance between superconductivity and antiferromagnetism and opening new ambient-pressure phase space in the study of heavy-fermion ground states.

PACS numbers: 71.27.+a, 74.70.Tx

The quarter century of work on heavy-fermion materials has developed a widespread belief that the rather rare and exotic superconductivity found among these intermetallics occurs only near a quantum-critical point where the Néel ($T_N$) or Curie ($T_C$) temperature of a magnetically ordered ground state approaches T = 0K. [1] With Ce-based materials, sufficiently high pressure can drive $T_N \rightarrow$ 0K, but there is no predictive understanding of whether and when superconductivity will emerge. We expect, conversely, an antiferromagnetic state lying in close proximity to any heavy-fermion superconducting ground state. In the CeCu$_2$Si$_2$ system, this is seen in detailed and intricate annealing/compositional/pressure studies [2]. Likewise, the set of iso-structural, iso-electronic compounds CeCoIn$_5$ ($T_c$=2.3K), CeRhIn$_5$ ($T_N$=3.8K) and CeIrIn$_5$ ($T_c$=0.4K) [3] have proven fertile for the study of heavy-fermion ground states and of the interplay between antiferromagnetism and superconductivity as Co or Ir is substituted systematically into the

Rh member [4]. With such a closely related sequence, there is the prospect for establishing exactly what separates the heavy-fermion antiferromagnetic ground state from the superconducting one; however, even given the same-column similarities of Co, Rh and Ir, their substitutional study has resisted understanding when superconductivity emerges. Similarly, electron doping the Ce115s by Sn substitutions for In has found only a monotonic suppression of superconductivity in $CeCo(In,Sn)_5$ [5] or antiferromagnetism in $CeRh(In,Sn)_5$ [6] without inducing a complementary ordered state.

We have found that Cd-doping Ce115 compounds provides a previously unappreciated perspective on this problem. A particularly clear example is our finding that $CeCoIn_5$, with the highest superconducting $T_c$ for a heavy-fermion material (2.3K), can be Cd-doped smoothly to an antiferromagnetic ground state, a state which itself can then be driven back to superconductivity with applied hydrostatic pressure. Cd-doping appears to mitigate problems encountered with previous substitutions, and shows, *inter-alia*, a surprising difference of behavior in the three compounds but with a systematic dependence from compound to compound in Cd concentration. We concentrate below on the $CeCo(In_{1-x}Cd_x)_5$ system, giving some comparison to the Cd-doped Ir and Rh members.

The simple tetragonal structure of $CeMIn_5$ (M=Co,Rh,Ir) of $HoCoGa_5$-type is a stacking along the c-axis of square planar layers of CeIn(1), $In(2)_2$, Co and $In(2)_2$. [7] Extensive Fermi surface studies [8] have found that the La-based Co, Rh and Ir 115's have very similar Fermi surfaces with two quasi-2D cylindrical hole-like pieces, plus smaller 3D electron surfaces. The Fermi-surface geometry of $CeRhIn_5$ is essentially the same, suggesting that the 4f-electron of Ce does not contribute appreciably to the Fermi volume. $CeCoIn_5$ and $CeIrIn_5$ have new deHaas-van Alphen frequencies, in addition to exhibiting frequencies which appear close to those of the quasi-2D cylindrical pieces found in the La compounds. It is interesting that CeRhIn$_5$ develops the CeCoIn$_5$-like Fermi-surface geometry above 2.3GPa [9], a pressure clearly greater than the pressure where superconductivity becomes the ground state and where Ce's 4f-electron appears to assume a more delocalized nature. [10] Characteristically, deHaas-van Alphen experiments are performed at high fields, typically of order 10 T or greater, corresponding approximately to an energy scale of 10 K which could obscure essential electronic fine structure. It may not be surprising, then, that band structure calculations find Fermi surfaces in reasonable agreement with experiment, but the predicted Fermi surfaces are the same for the three Ce115's. [11] This supports the view that the physics describing the interplay between competing heavy-fermion antiferromagnetic and superconducting ground states in the Ce115's lies entirely in the very low energy details of the electronic structure and, by inference, in some other heavy-fermion materials as well. The Cd-doped Ce115 experiments provide a new insight into this fundamental problem in correlated electron physics.

For these experiments, crystals were grown using a standard In-flux technique in which various amounts of Cd were added to the flux. Microprobe measurements of a series of CeCo(In$_{1-x}$Cd$_x$)$_5$ crystals found the In/Cd ratio to be very uniform, but with a Cd concentration consistently 10% of the nominal flux concentration across the entire range of flux compositions. An analysis of x-ray absorption fine structure measurements [12] on CeCo(In$_{1-x}$Sn$_x$)$_5$ samples showed that Sn occupied preferentially the In(1) position in the material. If this is so with Cd, then the Cd concentrations on the In(1) sites are approximately 50% that of the flux. This possibility is consistent with an estimate made in a preliminary NMR investigation from the intensity of the Cd signal on a nominal CeIr(In$_{0.90}$Cd$_{0.10}$)$_5$ material. [13] Although microprobe examination of the Ir and Rh materials has not been completed, we make the reasonable assumption that the Cd concentration in these crystals also is approximately 10% of that in the flux from which they were grown. Despite what we know from the above mentioned experiments of actual concentrations of Cd, nominal concentrations of x in $CeM(In_{1-x}Cd_x)_5$ (M=Co, Rh, Ir) will be stated throughout this Letter and labeled in the figures for clarity and continuity.

Samples were studied by specific heat, resistivity and magnetic susceptibility measurements performed in Quantum Design PPMS and MPMS apparatuses, respectively. Pressure-dependent resistance and *ac* susceptibility studies were carried out in a Be-Cu, clamp-type pressure cell containing a Teflon cup filled with silicone as the pressure-transmitting medium, samples and a small piece of Sn, whose inductively measured $T_c$ served as a manometer.

Figure 1 shows the evolution with increasing Cd content of the low temperature electronic specific heat of $CeCo(In_{1-x}Cd_x)_5$, $CeRh(In_{1-x}Cd_x)_5$ and $CeIr(In_{1-x}Cd_x)_5$ single crystals. These data, combined with magnetic susceptibility, resistivity and field-dependent specific heat measurements (not shown), reveal an unexpected response to very small Cd concentrations. In the Co-and Ir-115's, superconductivity gives way to antiferromagnetic order [14], which emerges first near nominal x=0.07 Cd doping, and with increasing Cd appears at temperatures exceeding that of undoped $CeRhIn_5$, which itself exhibits a non-monotonic variation of $T_N$ versus x. As shown in the inset of each panel, the magnetic entropy above 5-10 K is invariant to Cd concentrations where these changes in ground state occur.

Temperature-doping phase diagrams resulting from heat capacity measurements are summarized in Figure 2. In the case of Ir115 (panel (c)), we find only a magnetic ground state beyond the disappearance of superconductivity in the composition range near nominal CeIr $(In_{.95}Cd_{.05})_5$. This is the first example of magnetism appearing close to superconductivity in this compound, and interestingly, the magnetic ordering temperature of the Cd-doped Ir115 is the highest found among the Cd-doped Ce115s. This behavior can be compared with that found in the alloy system $Ce(Rh_{1-y}Ir_y)In_5$ [15] where substitutions are made away from the CeIn(1) plane. There a minimum in $T_c$ appears near $y = 0.9$, with superconductivity on both sides of the minimum and with some experimental evidence that the superconducting ground states may be different. [16] For $y \approx 0.6$, superconducting and antiferromagnetic ground states coexist. With $CeRhIn_5$ (panel (b)), there is a flat minimum in $T_N$ in the range of Cd concentrations x= 0.05-0.10. We speculate that this may be connected with an incommensurate to commensurate magnetic ordering that is induced in $CeRhIn_5$ when a magnetic field applied in the tetragonal basal plane. [17] With Sn having one more p-electron than In and Cd one less electron, it might be supposed that Sn-doping could have the opposite effect of Cd-doping, and if so, Sn-doping of $CeRhIn_5$ should lead to superconductivity, just as pressure does in this material. In contrast to this expectation, superconductivity is not found with Sn-doping; instead Sn only drives $T_N$ to 0K near the 2D percolation limit. [6] Likewise, Sn substitution for In in $CeCoIn_5$ does not enhance $T_c$ but suppresses it to zero with about 3-4% Sn and does not induce additional phase transitions. [5] Why this is so provides one of many avenues for further alloy studies in these materials on the border between magnetic order and exotic superconductivity.

Introduction of Cd into CeCoIn$_5$ (panel (a)) creates initially a two phase region above nominal x=0.075, where T$_N$>T$_c$, followed by only antiferromagnetism for x > 0.12. As with Cd-doping Ir115, the smooth evolution of T$_c$(x) and T$_N$(x) rules against real-space inhomogeneity. This phase diagram is remarkably similar to that found from studies of CeRhIn$_5$ under pressure. [10] A value of nominal x=0.15 corresponds roughly to CeRhIn$_5$ at a pressure of 0.9 GPa and removing a small number of electrons from CeCoIn$_5$ essentially reproduces the pressure-induced evolution of ground states in CeRhIn$_5$. This view is supported by pressure experiments on nominal x=0.10 and x=0.15 in CeCo(In$_{1-x}$Cd$_x$)$_5$, which are shown in Figure 3. Applying pressure to these doped materials accurately reverses what was seen with progressive Cd-doping, essentially the pressure behavior seen in undoped CeRhIn$_5$ [10]. Further, as shown in Figure 3, results for nominal x=0.10 and 0.15 superimpose with a rigid shift of their respective pressure axes by 0.7 GPa. A simple interpretation of this scaling implies that a nominal 5%-Cd concentration change in CeCoIn$_5$ acts like a negative pressure of 0.7 GPa, so that undoped CeCoIn$_5$ would correspond to a negative pressure of 2.1 GPa. Interestingly, this value is nearly identical to the positive pressure required to induce a CeCoIn$_5$-like Fermi surface in CeRhIn$_5$ [9]. Both the a- and c axis lattice parameters of the Cd-doped CeCoIn$_5$ are smaller than those of the undoped compound, so that chemical pressure effects of doping are in the direction of increased pressure and, consequently, the comparison to CeRhIn$_5$ suggests that electronic tuning, not chemical pressure, is most relevant. It should be noted that the superconducting transition is seen in resistivity for CeCo(In$_{1-x}$Cd$_x$)$_5$ (panel (a)) at temperatures higher in comparison to heat capacity for nominal concentrations of 0.075<x<0.15. A similar relationship between resistively and thermodynamically determined $T_c$s is observed in CeRhIn$_5$ over a range of pressures where magnetism and superconductivity coexist and $T_N(P) \geq T_c(P)$. [19]

The results found here are unlike what we are familiar with in heavy-fermion materials. The closest related example is CeCu$_2$(Si/Ge)$_2$ in which replacing Si with Ge expands the unit-cell volume and induces magnetic order that can be reversed with applied pressure [2]. Compared to Cd-doping CeCoIn$_5$, however, the disordered Si/Ge sublattice produces larger low temperature residual resistivities and unavoidably broadens details of the electronic structure. This is not the case here: the physics of the Cd-doped samples under pressure for different Cd concentration and compared to undoped CeRhIn$_5$ is very similar to that of the undoped materials. The specific heat anomalies in the Cd-doped materials remain relatively sharp in comparison to Sn-doping [5,6] and appear only somewhat affected by additional scattering coming from the addition of Cd.

One way that pressure may act is to shift energy bands relative to each other. Electron removal by Cd-doping also affects the Fermi surface by shifting the Fermi level, and, as implied from results from Figure 3, appears to work in the same manner as pressure with opposite sign. The experimental observations here suggest that the dominant effect of Cd substitution is to shift the Fermi energy without significant broadening of the energy features relevant to the low temperature physics. The low temperature specific heat data shown in Figure 1, along with the observation of how the entropy develops with temperature and doping, invite the speculation that both the antiferromagnetism and the superconductivity are Fermi surface instabilities. From this viewpoint the physics of CeCoIn$_5$ involves the interplay of superconductivity and antiferromagnetism instabilities on different parts of the Fermi surface: tuning the Fermi surface via pressure and/or doping shifts the balance favoring one or the other ground state. Dominance of one ground state over another can presumably be propagated through proximity effects. This is not a new idea in correlated electron physics, but it appears here naturally and in a way that has the promise of direct experimental access.

The experiments reported here on the heavily studied, archetypic heavy-fermion 115s appear to be telling us that the rich low temperature physics is being simply determined by electron count at the percent level. The ability to move easily between antiferromagnetic and superconducting ground states via small changes in the electron count at ambient pressure now gives the opportunity to address directly with a full arsenal of experimental tools central questions in heavy-fermion and more generally correlated electron materials, questions such as the relevance of quantum criticality, Kondo scale and crystal field effects to low temperature properties. Our suspicion is that what underlies the richness of the phenomena seen in the 115's is a complexity coming from Fermi surface detail, a detail in which Kondo and coherence scales are central and on which we now have a new handle.

We acknowledge useful discussion with L. P. Gor'kov, D. Pines, P. Schlottmann and F. Ronning. Work at Los Alamos was performed under the auspices of the US DOE/Office of Science. Work at UC Davis and UC Irvine has been supported by grant NSF-DMR 053360.

Figure 1 || (Color Online) Electronic specific heat ($C_{el}=C-C_{latt}$) divided by temperature for CeM(In$_{1-x}$Cd$_x$)$_5$ as a function of temperature. The lattice specific heat ($C_{latt}$) is not shown. Values of x represent the nominal Cd content of crystals, as explained in the text. (a) CeCo(In$_{1-x}$Cd$_x$)$_5$ (b) CeRh(In$_{1-x}$Cd$_x$)$_5$ (c) CeIr(In$_{1-x}$Cd$_x$)$_5$. The solid curve through the data for x=0.0375 in panel (c) is a fit to the form $C_{el}/T = A\ln T/T_0$, where A=240mJ/mole-K$^2$ and $T_0$=18K, for 0.4 T<4 K. Insets in each panel are plots of magnetic entropy, $S_{mag}$, obtained from the area under associated $C_{el}/T$ vs T curves.

Figure 2 || (Color Online) Dependence of superconducting transition temperatures $T_c$ and Néel temperatures $T_N$ on x, where x is the nominal Cd content of crystals. See text for details. (a) CeCo(In$_{1-x}$Cd$_x$)$_5$, (b) CeRh(In$_{1-x}$Cd$_x$)$_5$, and (c) CeIr(In$_{1-x}$Cd$_x$)$_5$. For (b), $T_N^A$ and $T_N^B$ are associated with different AFM phases as discussed in the text. $T_c$ and $T_N$ were extracted from specific heat (Figure 1) and confirmed with magnetic susceptibility measurements.

Figure 3 || (Color Online) Results of pressure studies on CeCo(In$_{1-x}$Cd$_x$)$_5$. (a) Pressure dependence of the Néel $T_N$, and superconducting transition temperature $T_c$ for CeRhIn$_5$ (black circles) [18] and CeCo(In$_{1-x}$Cd$_x$)$_5$ at nominal x=0.10 (blue squares) and 0.15 (red triangles). With CeRhIn$_5$ as the reference, a rigid shift of nominal x=0.15 data by +0.9 GPa and of nominal x=0.10 data by an additional +0.7 GPa (i.e., a total shift of 1.6 GPa relative to CeRhIn$_5$) superimpose all three sets of data. (b) Low temperature in-plane resistivity ($\rho_{ab}$) at atmospheric and 0.45 GPa pressures for nominal x=0.10. Bulk superconductivity is confirmed by simultaneous *ac* susceptibility measurement. (c) In-plane resistivity ($\rho_{ab}$) at

atmospheric and 0.95 GPa pressures for nominal x=0.15. Bulk superconductivity is confirmed by simultaneous *ac* susceptibility measurement.

Fig. 1 Pham, *et al.*

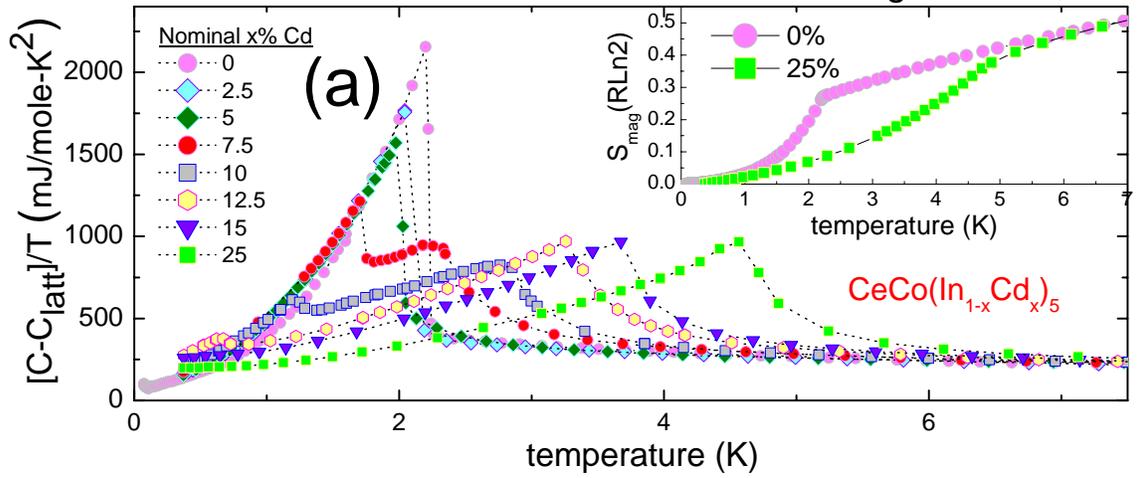
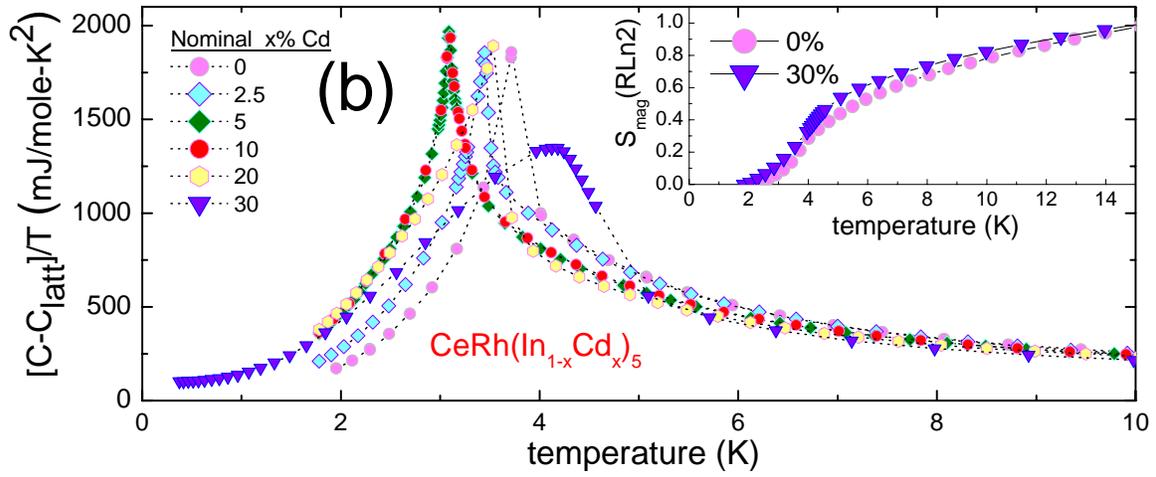
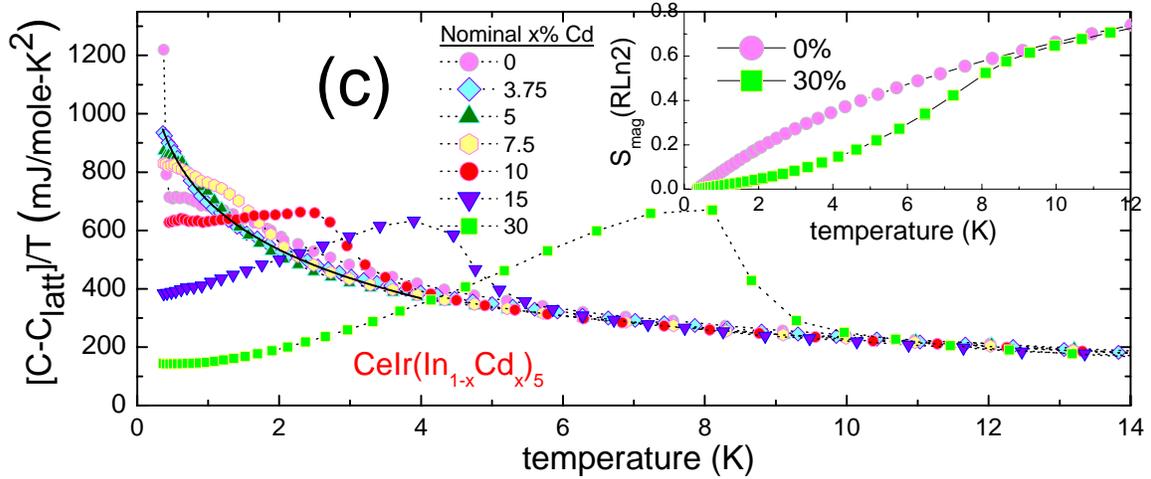

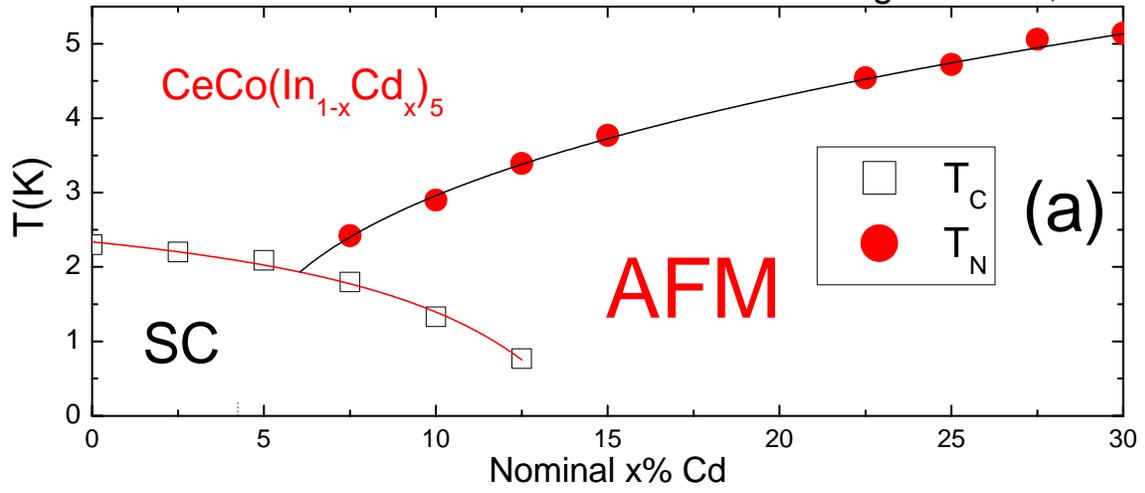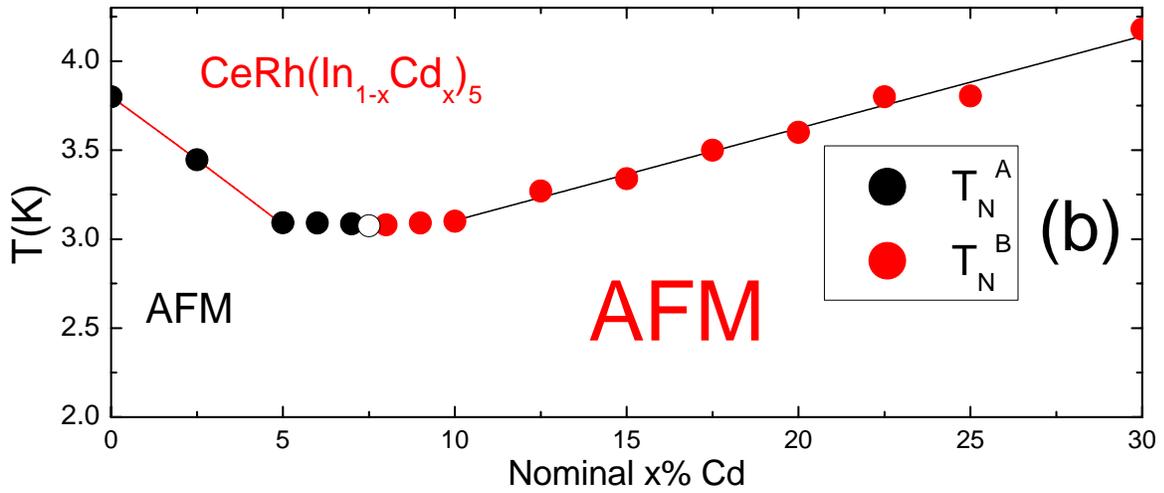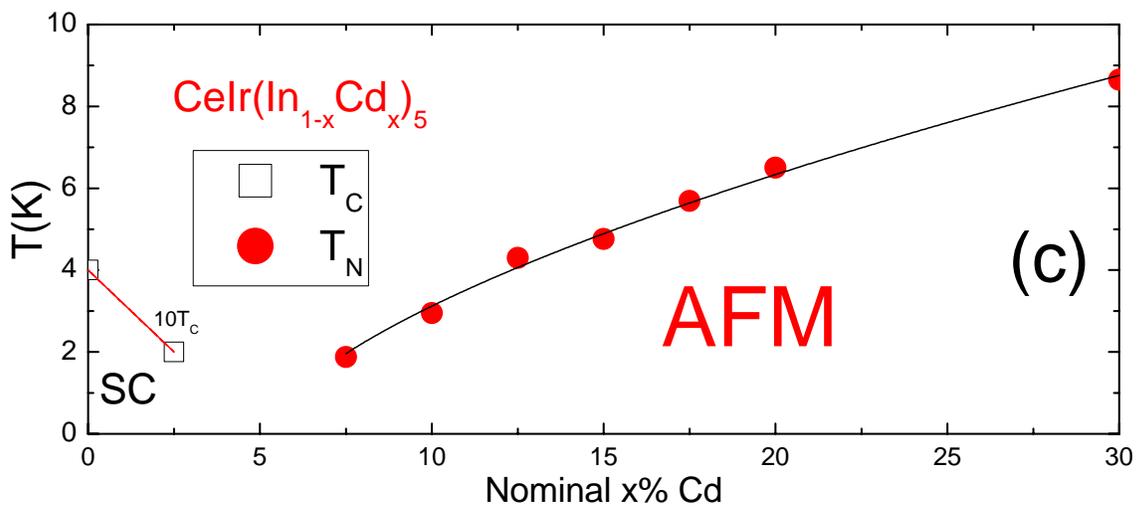

Fig. 2 Pham, et al.

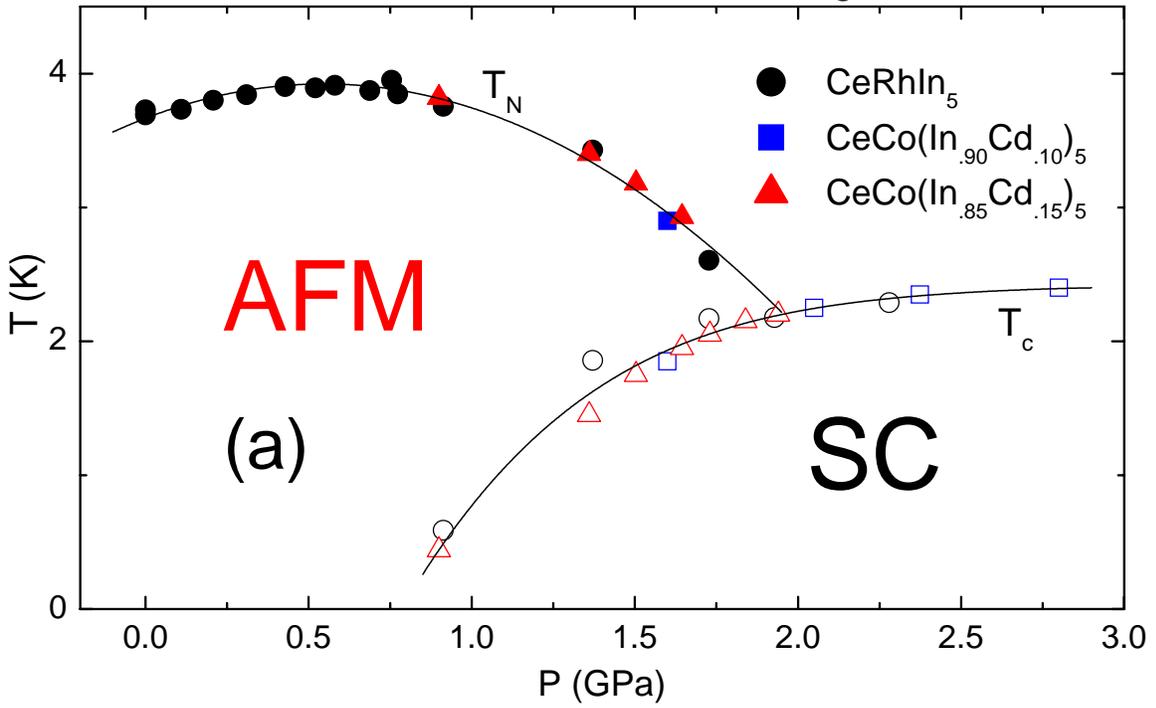
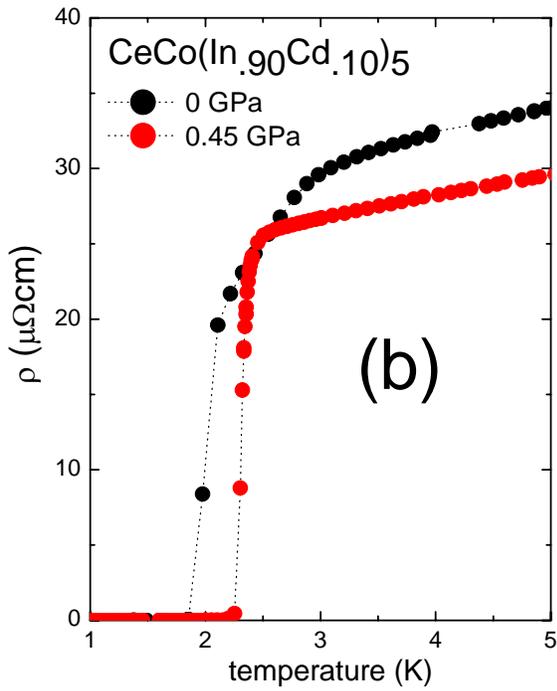
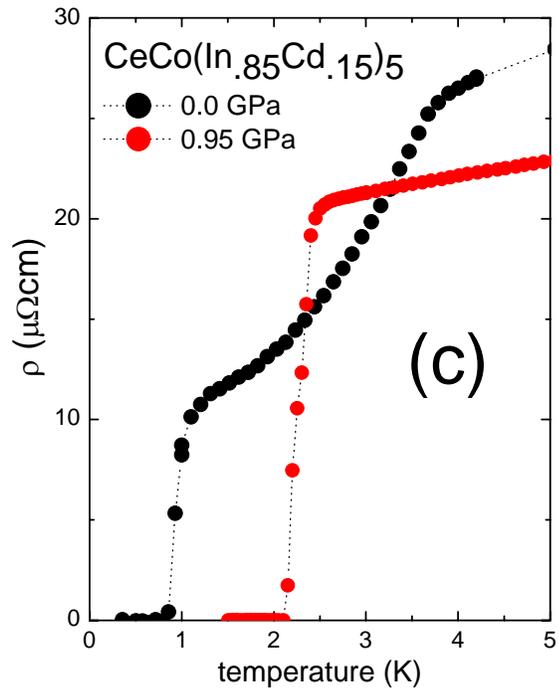